\documentclass[twocolumn, prd, showkeys, superscriptaddress]{revtex4}

\usepackage{textcomp}
\usepackage{makeidx}
\usepackage{amsmath}
\usepackage{subfigure}
\usepackage{amssymb}
\usepackage{hyperref}
\usepackage{graphicx}

\begin{document}

\title{Embedding of two de-Sitter branes in a generalized Randall Sundrum scenario.}

\author{Rodrigo Aros}
\email{raros@unab.cl}
\affiliation{Departamento de Ciencias F\'isicas, Universidad Andr\'es Bello, Av. Rep\'ublica 252, Santiago,Chile}
\author{Milko Estrada}
\email{milko.estrada@gmail.com}
\affiliation{Departamento de Ciencias F\'isicas, Universidad Andr\'es Bello, Av. Rep\'ublica 252, Santiago,Chile}
\affiliation{Universidad Internacional SEK, Fernando Manterola 0789, Providencia, Santiago, Chile}

\date{\today}

\begin{abstract}
In this work it is studied a generalization of Randall Sundrum model. It is obtained new behaviors of {\it warp factor}, constraints in that the strong brane has tension of positive sign and, new relations between the coupling constants $\kappa_4$, $\kappa_5$ and between the Higgs and Planck masses.
\end{abstract}

\maketitle

\section{Introduction}
The lack of a theory that unifies the four fundamental interactions is nowadays one of the most important issue of physics. In particular it is necessary to explain the huge difference between the values of the Higgs mass, $m_H \approx 1 TeV$, and the Planck mass $m_P \approx 10^{19}GeV$. This is (called) the \emph{hierarchy problem}. To address this problem in 1999 Lisa Randall and Raman Sundrum \cite{7} proposed a model (RS) including two statics 3-branes imbedded in AdS$_5$ space with tensions of equal magnitude but opposite signs and our universe corresponding to the positive tension brane. This space has also  $\mathbb{Z}_2$  symmetry. Using coordinates $x^M = (t,x,y,z,\phi)$, with $ \phi \in [-\pi,\pi]$, our universe is located at $\phi=0$ and the secondary brane, which is called the strong brane, at $\phi=\pi=-\pi$. In this model it is also proposed that Planck scales in four and five dimensions are of the same order, \emph{i.e.}, $M_{4} \approx M_{5}$.
The line element considered in this model is given by
\begin{equation}
ds_5^2=e^{-2kr|\phi|}  \eta_{uv}dx^udx^v + r^2 d\phi, \label{1}
\end{equation}
where $e^{-2kr|\phi|}$ is called warp factor. In this way the value of mass in the two different branes is related by $m_{\phi=0}=e^{-kr\pi}m_{\phi=\pi}$. This allows that  Planck mass at strong brane to be of same order of Higgs mass at our universe provided $e^{kr\pi} \approx 10^{15}$.

RS model however is not adequate to describe the current observations showing that our universe is under an accelerate expansion. Because of that in this work we study a model that  consider dS$_4$ brane-worlds imbedded in (A)dS$_5$. The line element is given in this case by
\begin{equation}\label{2}
ds^2_5 =  e^{-2A(\phi)}   \frac{L^2}{t^2} \eta_{uv}   dx^udx^v+r^2 d\phi^2 ,
\end{equation}
where $L$ is the radius dS$_4$.

In the next sections Einstein equations are solved yielding warp factors, tension of both branes and new relations between the coupling constants $\kappa_4$ and $\kappa_5$ and between Higgs and Planck masses. It will be analyzed the cases $\Lambda_{5D} = \pm \frac{6}{l^2}$ and $\Lambda_{5D} \to 0$.

\section{Embedding of two branes $dS_4$ in a space-time $(A)dS_5$.}

Let us consider the EH action coupled with two branes \cite{3}, \emph{i.e.},
\begin{align}
S= &\frac{1}{2 \kappa_5^2} \int dx^5 \sqrt{-g} \Big  [R^{(5D)}-2\Lambda_{5D} \Big ] - T_1  \int d^4 \sigma \sqrt{-h_0}  \nonumber \\
     &- T_2 \int d^4 \sigma \sqrt{-h_\pi}  .
\end{align}
where $\kappa_5$, $g$, $R^{5D}$ and $\Lambda_{5D}$ are the coupling constant, the metric determinant,  Ricci scalar and the cosmological constant $5D$  respectively. Here $T_1$, $T_2$, $h_0$ y $h_\pi$ are the tensions and the induced metrics on the weak and strong branes respectively. The equation of motion are
\begin{align}
G_{MN}^{(5D)}+\Lambda_{5D} g_{MN} =&-\kappa_5^2 \Big  [T_1\sqrt \frac{h^0}{g} h^0_{uv}\delta^u_M\delta^v_N \delta(\phi)  \nonumber \\
&+ T_2\sqrt \frac{h^{\pi}}{g} h^{\pi}_{uv}\delta^u_M\delta^v_N \delta(\phi-\pi) \Big ]   .   \label{4}
\end{align}

The $(\phi, \phi)$ component of Eq.\eqref{4},
\begin{align}
(A')^2  =  r^2 \Big ( \frac{e^{2A(\phi)} }{L^2} \mp \frac{1}{l^2} \Big ), \label{5}
\end{align}
takes values $\mp$ for $\Lambda_{5D} = \pm \frac{6}{l^2}$. The $(u,v)$ component is
\begin{align}
&G_{uv}^{(4D)}-\frac{3}{r^2}g_{uv}(\sigma^u)e^{-2A(\phi)} \Big (A''-2(A')^2 \Big ) \nonumber \\
&\pm \frac{6}{l^2}  g_{uv}(\sigma^u)e^{-2A(\phi)} = \nonumber \\
& -\kappa_5^2 \Big [T_1\sqrt \frac{h^0}{g} h^0_{uv} \delta(\phi) + T_2\sqrt \frac{h^{\pi}}{g} h^{\pi}_{uv} \delta(\phi-\pi) \Big ] , \label{6}
\end{align}
where $g_{uv}(\sigma^u)=\frac{L^2}{t^2}\eta_{uv}$. Replacing Eq.\eqref{5} in Eq.\eqref{6}, for the values of $\phi \neq 0 \neq \pi$, yields
\begin{equation}\label{FourDEE}
G_{uv}^{(4D)} = - \frac{3}{L^2} g_{uv}(\sigma^u).
\end{equation}
The last condition is satisfied for the space $dS_4$.  Therefore a factor $A(\phi)$ which is solution of the $(\phi, \phi)$ Einstein equation, is also also a solution of the $(u,v)$ component for values $ \phi \neq 0 \neq \pi$. The induced metrics on the weak and strong branes are respectively
\begin{align}
h_{uv}^{0}=e^{-2A(0)}g_{uv}(\sigma^u) \textrm{ and }\nonumber \\
h_{uv}^{\pi}=e^{-2A(\pi)}g_{uv}(\sigma^u). \label{7}
\end{align}

Replacing the Eqs. \eqref{5}, \eqref{FourDEE} and \eqref{7} in Eqs.\eqref{6} yields
\begin{equation}\label{8}
  \frac{3}{L^2}e^{2A(\phi)}    -\frac{3}{r^2} A''  =
- \kappa_5^2 \Big [   \frac{T_1}{r} \delta(\phi) +  \frac{T_2}{r}     \delta(\phi-\pi) \Big ].
\end{equation}
This yields the tensions
\begin{align}
\frac{3}{r \kappa_5^2} A' \Big |^\epsilon_{-\epsilon} &= T_1, \label{9}  \\
\frac{3}{r \kappa_5^2} A' \Big |^{-\pi+\epsilon}_{\pi-\epsilon} &= T_2 . \label{10}
\end{align}

The tension of a brane acts as a cosmological constant, which on our brane universe is associated with a energy density of positive sign and a negative pressure. Therefore, solutions that yield $T_1<0$ are not to be considered.

\section{four dimensional effective theory}
Since the extra dimension, said $\phi$, can not be detected by current experiments it is interesting to consider the form adopted by EH action in five dimensions. This is given by
\begin{equation}\label{11}
S_{eff} = \frac{r}{\kappa_5^2}  \int^\pi_{-\pi} d\phi e^{-2A} \int d^4x \sqrt{g^{(4D)}}  R^{(4D)} \subset S.
\end{equation}
By comparing this equation with the four dimensional EH action it is obtained the relation
\begin{equation}\label{12}
\frac{r}{\kappa_5^2}  \int^\pi_{-\pi} d\phi e^{-2A} = \frac{1}{\kappa_4^2} .
\end{equation}
where the coupling constants $\kappa_4$ and $\kappa_5$ are related with the Planck scales $M_4$ and $M_5$ through the relation
\[
\kappa^2_{4+n}= \frac{8\pi}{M_{4+n}^{2+n}}.
\]

To proceed is necessary to fix some parameters. Recalling \cite{7} it is convenient to fix the values of $\kappa_4$ and $\kappa_5$ such that $M_4 \approx M_5$.

To propose new relations between the Higgs and Planck mass it will be proceed \emph{\`a la  } Randall and Sundrum by considering the action for a Higgs field on the strong brane. This is
\begin{equation}
S_\pi \subset \int d^4x \sqrt{-h_\pi} \Big [ h_\pi^{uv}D_u H^{\dag}D_vH- \lambda ( |H|^2-v_\pi^2)^2 \Big],
\end{equation}
where $H$ stands for the Higgs field. Here $v_\pi$  is the Higgs vacuum expectation value, $\lambda$ is the coupling constant and $D_u$ is the gauge covariant derivative.  By normalizing $H \to e^{A(\pi)}H$ equation \eqref{7} implies
\begin{align}
S_{eff} \subset& \int d^4x \sqrt{-g(\sigma^u)} \Big [ g^{uv}(\sigma^u)D_u H^{\dag}D_vH \nonumber \\
&- \lambda ( |H|^2-e^{-2A(\pi)}v_\pi^2)^2 \Big],
\end{align}
where the vacuum expectation value is suppressed. This yields $v \equiv e^{-A(\pi)}v_\pi$. Any mass parameter at weak brane correspond to the physical mass defined by
\begin{equation}\label{15}
m_{\phi=0} \equiv e^{-A(\pi)}m_{\phi=\pi}.
\end{equation}

In the rest of paper we will study in different scenarios the values of $r$ that produce a mass of order of Higgs in our brane universe considering a mass of order of Planck at the strong brane.

\subsection{Case $\Lambda_{5D} \to 0$}

The line element is
\begin{equation}
ds^2=   \Big (|\phi| - C \Big )^2    \frac{r^2}{t^2} \eta_{uv}dx^udx^v +  r^2 d \phi ^2 , \label{16}
\end{equation}
where it can be noticed that there is not an exponential dependence. Therefore, the scaling differs from Randall Sundrum model along the fifth dimension.

By replacing the factor $A(\phi)$ of equation \eqref{16} in equations \eqref{9} and \eqref{10}, the tensions are
\begin{align}
&T_1=\frac{6}{\kappa_5^2 r C} ,\label{17} \\
&T_2=\frac{6}{\kappa_5^2 r(\pi  - C)},  \label{18}
\end{align}
where the constant $C >0$ and therefore $T_{1}>0$. From equation \eqref{12} is obtained
\begin{equation}
\frac{2}{3} \frac{\pi r^3}{L^2 \kappa_5^2} \Big ( \pi^2  - 3 \pi C + 3 C^2 \Big ) = \frac{1}{\kappa_4^2}, \label{19}
\end{equation}
where it can be observed that the $r$ dependence is no longer weak and thus differs from Randall Sundrum model. On the other hand, from equation \eqref{15} arises the relation between the mass  $m_{\phi=0}$ y $m_{\phi=\pi}$ given by
\begin{equation}
m_{\phi=0}= \Bigg | \frac{r}{L} \Big ( \pi - C \Big) \Bigg | m_{\phi=\pi}.    \label{20}
\end{equation}

First at all, we will study the case where $C=\frac{E}{r}$. In this case the line element \eqref{16} represents a Minkowski space at $\phi=0$ and at a time of order of the current age of our universe, $t = E \approx 10^{61} t_p \approx 10^{61} l_p$  (where $l_p$ and $t_p$ are the lenght and time of Planck, respectively) \cite{27}. With this value of $C$, the tensions are:
\begin{align}
&T_1=\frac{6}{\kappa_5^2 E} ,\label{21} \\
&T_2=\frac{6}{\kappa_5^2(\pi r - E)}.  \label{22}
\end{align}
It is worth to notice that $T_1>0$ but $T_2<0$ or $T_2>0$ provided $r<\frac{E}{\pi}$ or $r>\frac{E}{\pi}$. If $E \approx 10^{61} l_p$ the last constraint is uninteresting, since, large values for $r$ would imply deviations from Newtonian gravity at solar system scales \cite{6}. It is interesting to observe that at the limit $r \to 0$ the Randall and Sundrum model is reproduced, \emph{i.e.} $T_1=-T_2$.

By substituting $C=\frac{E}{r}$ in equation \eqref{19}, we obtain the relation between $\kappa_4$ and $\kappa_5$:
\begin{equation}
\frac{2}{3} \frac{\pi r}{L^2 \kappa_5^2} \Big ( \pi^2 r^2 - 3 \pi r E + 3 E^2 \Big ) = \frac{1}{\kappa_4^2}, \label{23}
\end{equation}
where, using $M_5 \approx M_4 \approx M_{Planck}=10^{19}GeV$ yields, $\kappa_4^2 \approx 10^{-38} GeV^{-2}$ and $\kappa_5^2 \approx 10^{-57} GeV^{-3}$. On the other hand, considering $\Lambda_{4D} = \frac{3}{L^2} \approx 10^{-122}l_p^{-2}$ it turns out that $L \approx 10^{61}l_p$. Finally, using a value comparable to the age of universe $E=2.820947918 \cdot 10^{60} l_p$, we find a radius of order of $2 l_p$.

With these values for $E$ and $L$ in the equation \eqref{20} implies a separation between branes $\pi r$ of approximately $10^{45} l_p + E$ to establish a mass of order of Higgs on our brane universe. This, unfortunately, is not a reasonable value from physical point of view.

Because of that one can consider that $C$ is not of the same order of $\frac{E}{r}$. For instance, if $C$ is a constant of value $C \approx 5 \cdot 10^{44}$ then a mass of order of Higgs arises on our brane universe for values of $L \approx 10^{61}l_p$ and $r \approx 2 l_p$.  Furthermore, the tensions $T_1$ and $T_2$ have positive and negative signs, respectively. Equation \eqref{19} gives in this cases that $10^{-10}M_5 \approx M_4$.

Despite the fact that in this last case we do not obtain that  $M_4 \approx M_5$, a null five dimensional cosmological constant could be of physical interesting, since at the above paragraph we obtain a mass of order of Higgs on our brane universe, which has tension $>0$.

\subsection{Case $\Lambda_{5D} < 0$}

In this case is necessary to consider the line element
\begin{equation}\label{24}
ds^2= l^2 \sinh^2 \Big (  \frac{r}{l}| \phi| -  C  \Big ) \frac{1}{t^2}  \eta_{uv}dx^udx^v + r^2 d \phi ^2,
\end{equation}
which for $r=1$ is similar to reference \cite{8}. In order to recover a Minkowski space at a time $t=E$ and $\phi=0$  it is necessary to fix $ C =  \sinh ^{-1} ( \frac{E}{l})$.  In this case the tensions are given by
\begin{align}
&T_1 = \frac{6}{\kappa_5^2 E} \cosh \bigg (  \sinh ^{-1} \Big (\frac{E}{l} \Big ) \bigg ), \\
&T_2=\frac{6}{\kappa_5^2 l } \coth \Big (\frac{r \pi}{l} - \sinh ^{-1} \big (\frac{E}{l} \big ) \Big  ).
\end{align}
It is worth to notice that $T_1>0$ but $T_2$ could be positive or negative depending on the $r$.

For $l \approx l_p$ \footnote{$\Lambda_{5D} \approx M_4^2$, same to Randall and Sundrum model, how it is indicated at the reference ~\cite{14}} and $E \approx 10^{61}l_p$, $T_2<0$ for  $r<44.91 l_p$ and $T_2>0$ for $r>44.91 l_p$. Therefore a value of $r$ near and greater than $44.91 l_p$, which is $\approx 1.5$ times bigger than the radio of Randall and Sundrum model could be of physical interest. In this last case the model has two branes of positive sign.

Unfortunately the relation between constants  $\kappa_4$ and $\kappa_5$ in this case is highly non linear. This is given by
\begin{align}\label{27}
&\frac{l}{\kappa_5^2} (\frac{l}{L})^2 \bigg [  \cosh \Big ( \frac{r}{l} \pi -  \sinh ^{-1} (\frac{E}{l}) \Big ) \sinh \Big (\frac{r}{l} \pi -  \sinh ^{-1} (\frac{E}{l}) \Big  ) \nonumber \\
& -\frac{r}{l}  \pi +  \frac{E}{l}   \cosh \Big (  \sinh ^{-1} (\frac{E}{l}) \Big )  \bigg ] = \frac{1}{\kappa_4^2}.
\end{align}
In this relation, as in Eq. \eqref{23}, it is observed a dependence on $r,L$ and $E$. In this equation is difficult to adjust all constants, due to the number of terms and the sensibility of equation under small changes on the values of the constants.

The mass relation is
\begin{equation}\label{28}
m_{\phi=0}= \Bigg | \Big (\frac{l}{L} \Big ) \sinh \Big (  \frac{r}{l} \pi - C \Big ) \Bigg | m_{\phi=\pi},
\end{equation}
where, again using $l \approx l_p$ and $E \approx 10^{61}l_p$, it is required a radius $ r=78.133 l_p$ to produce a Higgs mass on our brane universe. This radius is $ \approx 2.6$ times bigger than Randall and Sundrum model one.  It is remarkable that with this value of $C$ it can be reproduced a Higgs mass on our brane universe.

\subsection{Case $\Lambda_{5D} > 0$}
In this case is worth considering a line element of the form
\begin{equation} \label{29}
ds^2=  \sin^2 \Big (  \frac{r}{l}| \phi| \pm  C \Big ) \frac{l^2}{t^2}  \eta_{uv}dx^udx^v + r^2 d \phi ^2.
\end{equation}
The tension of branes are respectively
\begin{align}
&T_1 =\mp \frac{6}{\kappa_5^2 l} \cot \Big (  C \Big ) , \label{30}  \\
&T_2=\frac{6}{\kappa_5^2 l } \cot \Big (\frac{r \pi}{l} \pm  C  \Big ). \label{31}
\end{align}
The relation between  $\kappa_4$ and $\kappa_5$ is given by
\begin{align}
&\frac{l^2}{L^2 \kappa_5^2} \bigg ( \pm 2 l \cos (C) \sin (C) - 2 l \cos \Big (\frac{\pi r}{l} \Big) \sin \Big  (\frac{\pi r}{l}\Big ) \cos^2 (C) \nonumber \\
&\mp 2 l \cos^2 \Big  (\frac{\pi r}{l}\Big ) \cos(C) \sin (C)+ l \cos \Big (\frac{\pi r}{l} \Big ) \sin \Big (\frac{\pi r}{l} \Big ) + \pi r \bigg ) \nonumber \\
&=\frac{1}{\kappa_4}. \label{32}
\end{align}
Finally, this yields a relation between the masses
\begin{equation}\label{33}
m_{\phi=0}= \Bigg | \Big (\frac{l}{L} \Big ) \sin \Big (  \frac{r}{l} \pi \pm  C \Big ) \Bigg | m_{\phi=\pi}.
\end{equation}

For $C=\sin ^{-1} \Big ( \frac{E}{l}\Big )$ the space at the brane is Minkowski at $\phi=0$ and $t=E$. Since there is not constraints for the value of cosmological constant in five dimensions it will be consider $|\frac{E}{l}|  \le 1 $. Furthermore, due to that $E$ and $l$ are positive constants, $C$ is located in the first two quadrants.

By taking $E \approx 10^{61}l_p$, $l \approx \sqrt{2}E$ and $r \approx l_p$, the constant $C$ can be determined depending on the $\pm$ in Eq.(\ref{32}). For $C=C^{+} = \frac{\pi}{4}$  and $C=C^{-}=\frac{3}{4}\pi$.  Furthermore, and independently of the sign $\pm$, $T_1>0$ and $T_2<0$  respectively.

Considering the values of  $E$, $l$ and $r$, the relation \eqref{32}, it is interesting to notice that it can be imposed that $ \mp l \pm l \cos ^2 (\frac{\pi r}{l}) \approx 0$. In this case
\begin{equation}
\frac{r}{\kappa_5^2} \Big (\frac{l}{L} \Big)^2  \pi = \frac{1}{\kappa_4^2},  \label{34}
\end{equation}
where the constants are adjusted such that $M_4 \approx M_5$ and $L \approx 10^{61}l_p$. Finally for the masses in relation \eqref{33}, it is not obtained a mass of order of Higgs.

Now, using a constant $C= \mp \frac{\pi r}{l} \mp \sin ^{-1} \big (\frac{10^{45}l_p}{l} \big )$ on equation \eqref{33}, we can obtain a mass of order of Higgs with the values $r \approx 6 l_p$, $l \approx 10^{61}l_p$ and $L \approx 10^{61} l_p$. With this values the equation \eqref{32} yields the relation $M_4 \approx 3.42 M_5$, where  $M_4=M_{planck}$ and, it is obtained a model where, at the equations \eqref{30} and \eqref{31}, $T_1$ is positive and $T_2=-T_1$.

\section{Conclusions}
It have been established new constraints for the values of compactification radius $r$ at the scenarios $\Lambda_{5D} \pm \frac{6}{l^2}$ and $\Lambda_{5D} \to 0$. These constraints allow to consider models where the tension of our brane is positive, the Planck constants $4D$  and $5D$ are approximately of same order, and a mass of order of Higgs on our brane universe is recovered.

A very interesting result is obtained at the case $\Lambda_{5D}>0$  with  $C= \mp \frac{\pi r}{l} \mp \sin ^{-1} \big
(\frac{10^{45}l_p}{l} \big )$, \emph{, since all assumptions described above are derived}.

Other interesting result is obtained for $\Lambda_{5D}<0$ where for a reasonable radius of compactification one recovers a mass of order of Higgs with $E$ of order of age of universe and two branes have positive tensions.

\end{document}